\documentclass[iop,twoside]{emulateapj}
\usepackage{graphicx,psfig,amssymb,float,natbib,rotating,txfonts}

\begin{document}

\shorttitle{The near-infrared counterpart to the neutron-star low-mass X-ray binary GX 3+1}
\shortauthors{van den Berg et al.}

\title{Discovery of the near-infrared counterpart to the luminous neutron-star
  low-mass X-ray binary GX 3+1$^{*}$} \thanks{This paper
  includes data gathered with the 6.5 meter Magellan Telescopes
  located at Las Campanas Observatory, Chile.}

\author{Maureen van den Berg\altaffilmark{1,2}, Jeroen Homan\altaffilmark{3},  Joel K. Fridriksson\altaffilmark{1}, and Manuel Linares\altaffilmark{4,5}}
\affil{\altaffilmark{1}Anton Pannekoek Institute for Astronomy,
  University of Amsterdam, Science Park 904, 1098 XH Amsterdam, The
  Netherlands; M.C.vandenBerg@uva.nl}
\affil{\altaffilmark{2}Harvard-Smithsonian Center for Astrophysics,
    60 Garden Street, Cambridge, MA 02138, USA}
\affil{\altaffilmark{3}Massachusetts Institute of Technology, Kavli
  Institute for Astrophysics and Space Research, 70 Vassar Street,
  Cambridge, MA 02139, USA}
\affil{\altaffilmark{4}Instituto de Astrof\'{i}sica de Canarias (IAC), V\'{i}a L\'{a}ctea s/n, La Laguna, E-38205, S/C de Tenerife, Spain}
\affil{\altaffilmark{5} Universidad de La Laguna, Departamento de Astrof{\'i}sica, E-38206 La Laguna, Tenerife, Spain}

\begin{abstract}
Using the High Resolution Camera onboard the {\em Chandra} X-ray
Observatory, we have measured an accurate position for the bright
persistent neutron-star X-ray binary and atoll source GX\,3+1. At a
location that is consistent with this new position we have discovered
the near-infrared (NIR) counterpart to GX\,3+1 in images taken with
the PANIC and FourStar cameras on the Magellan Baade Telescope. The
identification of this $K_s=15.8\pm0.1$ mag star as the counterpart is
based on the presence of a Br\,$\gamma$ emission line in a NIR
spectrum taken with the FIRE spectrograph on the Baade Telescope. The
absolute magnitude derived from the best available distance estimate
to GX\,3+1 indicates that the mass donor in the system is not a
late-type giant. We find that the NIR light in GX\,3+1 is likely
dominated by the contribution from a heated outer accretion disk. This
is similar to what has been found for the NIR flux from the brighter
class of Z sources, but unlike the behavior of atolls fainter
($L_X\approx 10^{36-37}$ erg s$^{-1}$) than GX\,3+1, where
optically-thin synchrotron emission from a jet probably dominates the
NIR flux.
\end{abstract}

\keywords{accretion, accretion disks; X-rays: binaries; binaries:
  close; stars: individual (GX\,3+1)}

\section{Introduction} \label{sec_intro}

Low-mass X-ray binaries (LMXBs) are systems in which a neutron star or
black hole accretes matter from a low-mass secondary ($M \lesssim 1
\,M_\odot$). Most are bright in X-rays only occasionally, but some
have been very X-ray luminous since their discovery.  Neutron-star
low-mass X-ray binaries (NS-LMXBs) with relatively weak magnetic
fields show a wide variety of correlated spectral and variability
behavior in X-rays. Based on this, two main sub-classes are recognized
\citep{hasivand89}: the so-called ``Z'' sources, with luminosities
close to or above the Eddington luminosity ($L_{\rm Edd}$), and the
``atoll'' sources, with luminosities up to $\sim$0.5 $L_{\rm Edd}$;
see \cite{homaea10} for an overview.

To explain the high mass-accretion rates ($\sim$$10^{-9}$--$10^{-8}
M_{\odot}$ yr$^{-1}$) that are implied by their X-ray luminosities, it
has been suggested that the mass donors in the most luminous ($L_X
\gtrsim 1\times10^{37}$ erg s$^{-1}$) NS-LMXBs are evolved stars
\citep{webbea83,taam83}. The evolutionary expansion of (sub-)giant
companions can drive mass transfer rates in excess of those that occur
in systems with main-sequence donors, where mass-transfer is driven by
the loss of angular momentum. The evolutionary stage of the donor also
affects the duration of the active X-ray--binary phase in the lifetime
of an LMXB, and might shape the LMXB X-ray luminosity function
(XLF). \cite{revnea11} suggested that the break around $\log L_X ({\rm
  erg~s}^{-1})\approx10^{37.3}$ in the LMXB XLF of our own Galaxy and
nearby galaxies \citep{gilv04} separates systems with main-sequence
companions from those with giant donors whose shorter active lifetimes
would explain the steepening of the XLF. Their argument uses orbital
period as a proxy of donor size, and observations of a handful of
systems confirm that this is a valid assumption in those
cases. However, for about half of the sources with $L_X>10^{37.3}$ erg
s$^{-1}$ an orbital period or donor classification is not available.

Optical and near-infrared (OIR) studies can potentially clarify some
of these issues. They shed light on the structure of the various
accretion flow components, such as disks and jets \citep{russea07},
and can provide information on the binary parameters, such as orbital
period and properties of the mass donor \citep[e.g.~][]{bandea99}.
Given that the population of bright NS-LMXBs is concentrated towards
the heavily-obscured Galactic bulge, the low extinction in the
near-infrared (NIR)---$A_K\approx0.1\,A_V$---makes it the preferred
band to carry out such studies in that region.

With a luminosity of $(2-4)\times10^{37}$ erg s$^{-1}$ \citep[2--10
  keV; ][]{denhea03} GX\,3+1 is one of the eleven most luminous and
persistently bright Galactic NS-LMXBs.  Ever since its discovery in
1964 \citep{bobych1965} observations of GX\,3+1 have found it to be a
bright X-ray source. The detection of X-ray bursts \citep{mamiin1983}
indicates it is a neutron-star system. The best distance estimate of
$\sim$6.1 kpc is derived from the properties of a radius-expansion
burst assuming an Eddington limit that is appropriate for a
hydrogen-poor atmosphere \citep{kuva2000,denhea03}.  The resulting
maximum persistent bolometric luminosity is $\sim$$6\times10^{37}$ erg
s$^{-1}$ or $\sim$0.3 $L_{\rm Edd}$. Based on its spectral and
variability properties GX\,3+1 is classified as an atoll source
\citep{hasivand89}. Similar to the bright atolls GX\,9+1 and GX\,9+9,
GX\,3+1 shows strong long-term X-ray flux modulations, which have a
time scale of $\sim$6 yr \citep{koch2010,duraea10}.  \citet{koch2010}
suggested these modulations are the result of variations in the
mass-transfer rate due to a solar-type magnetic-activity cycle in the
donor star. However, since the long-term X-ray light curve of GX\,3+1
covers only $\sim$2.5 cycles of the brightness modulations, it is
unclear if they are quasi-periodic or have a more random nature and
are simply a manifestation of very-low--frequency noise. The lack of
an accurate position for GX\,3+1 has hampered the search for a
counterpart, and so far the orbital period of GX\,3+1 remains unknown.

\cite{naylea91} were the first to look for a NIR counterpart to
GX\,3+1. They identified one candidate inside the region defined by
the {\it Einstein} error circle and the lunar-occultation error
box. More recently \cite{zolorevn11} repeated the search using an
additional position constraint imposed by a {\em Chandra} observation
performed in continuous-clocking mode.  In the highly-elongated region
of about $0\farcs5\times2$\arcsec\,they identified two possible
counterparts, viz.~the one originally found by \cite{naylea91} and a
much fainter source.

Here we present the results of our own search for GX\,3+1 in the NIR.
Combining a newly-determined accurate X-ray position with NIR imaging
and spectroscopy, we have uncovered the true counterpart of GX\,3+1
five decades after its discovery in X-rays. We describe the X-ray and
NIR observations in Section \ref{sec_obs}, and the identification of
the counterpart in Section \ref{sec_id}. In Section \ref{sec_disc} we
discuss the origin of the NIR emission in GX\,3+1, and make a
comparison with other NS-LMXBs.

\begin{figure}
\centerline{\includegraphics[angle=-90,width=8.3cm]{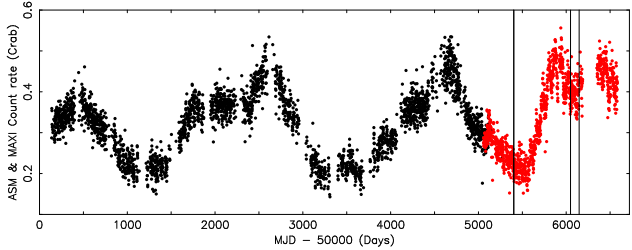}}
\caption{Long-term 2--10 keV {\em RXTE}/ASM (black points) and {\em
    MAXI} (red points) light curve of GX\,3+1. Count rates were
  normalized to those of the Crab. Vertical lines indicate the times
  of our Magellan imaging observations. \label{fig_xlightc}}
\end{figure}

\section{Observations and data reduction} \label{sec_obs}

\subsection{X-rays}

\subsubsection{Chandra HRC observation} \label{sec_hrc}

We observed GX\,3+1 with the {\em Chandra} High Resolution Camera
imaging detector (HRC-I; \citealt{zombea95}) for 1.2 ks, starting on
2012 February 12 at 19:22:43 UTC. The High Energy Transmission Grating
was inserted to lower the count rate and to limit distortions in the
image that could adversely affect the image reconstruction. The data
were analyzed using CIAO 4.5 \citep{frusea06}. We ran the {\tt
  wavdetect} tool on a 200$\times$200 pixel image, centered on
GX\,3+1. This yielded the following source position (R.A., Dec.):
17$^{\rm h}$47$^{\rm m}$56.077$^{\rm s}$,
$-$26$^\circ$33\arcmin49.48\arcsec~(J2000); the formal {\em wavdetect}
errors are negligible ($<$0\farcs01 in each coordinate). About
$2.1\times10^4$ photons were detected in a 2\arcsec-radius circle
centered around this position. No other X-ray sources were detected in
the field, which made it impossible to improve the absolute astrometry
of the image using X-ray sources with known accurate (radio or
optical/NIR) counterpart positions. Therefore, the positional accuracy
is limited by {\it Chandra}'s absolute astrometric calibration, which
for the HRC-I is $\sim$0\farcs54 (90\%
confidence)\footnote{http://cxc.cfa.harvard.edu/cal/ASPECT/celmon/}.

\subsubsection{{\it RXTE}/ASM and {\it MAXI} light curves}

We used public data from the All Sky Monitor (ASM) aboard the {\em
  Rossi X-ray Timing Explorer} \citep[{\em RXTE};][]{lebrcu1996} and
from the Monitor of All-sky X-ray Image ({\em MAXI}) camera
\citep{makaue2009} to construct a long-term 2--10 keV light curve of
GX\,3+1 from one-day average count rates ($ctr$). Data points with
large uncertainties ($ctr/\sigma_{ctr}<20$, with $\sigma_{ctr}$ the
standard deviation in $ctr$) were removed; for the {\em MAXI} data we
also removed a few-day--long flare around MJD 55916 whose origin and
relation to GX\,3+1 are unclear, and data between MJD 56187 and MJD
56351, which were affected by an outburst from a transient in the
globular cluster Terzan 5. The count rates from both instruments were
normalized to those of the Crab. The resulting light curve is shown in
Figure \ref{fig_xlightc}. Strong long-term modulations with
time-scales of several years are clearly visible. The vertical lines
in Figure \ref{fig_xlightc} indicate the times of our NIR imaging
observations (see next section); the first two were only separated by
about one day and appear as a single line.

\subsection{Near-infrared} 

\begin{table}
\caption{Magellan NIR observations of the GX\,3+1 field \label{tab_log}}
\begin{center}
\begin{tabular}{cccccc}
\hline
\hline
Epoch & Date & MJD\tablenotemark{a} & Instrument & T$_{\rm exp}$  & seeing \\
      & (UT) & (UT)                 &            & (s)          & (\arcsec) \\
\hline
1 & 2010 Jul 25 & 55402.0304 & PANIC     & 540 & 1.1 \\
2 & 2010 Jul 25 & 55402.9887 & PANIC     & 540 & 0.6 \\ 
3 & 2012 May 2  & 56049.3916 & FourStar  & 594 & 0.4 \\
4 & 2012 May 3  & 56050.3842 & FIRE      & 571 & 0.6 \\
5 & 2012 Aug 8  & 56147.1026 & FourStar  & 705 & 1.1 \\
\hline 
\end{tabular}
\end{center}
$^{a}$Modified Julian Date at the midpoint of the observation.
\end{table}

\subsubsection{PANIC imaging} \label{sec_panic}

A log of all our NIR observations is given in Table~\ref{tab_log}.  We
observed the field of GX\,3+1 with the Persson's Auxiliary Nasmyth
Infrared Camera (PANIC; \citealt{martea04}) on the 6.5-m Magellan
Baade telescope in Las Campanas, Chile. Mounted on the Baade, PANIC's
1024$\times$1024 pixel$^{2}$ HgCdTe detector has a 0\farcs127
pixel$^{-1}$ plate scale, and a field of view of about
2$\arcmin\times$2$\arcmin$. Observations were obtained through the
$K_s$ filter (with a $\sim$1.95--2.35 $\micron$ bandpass) on two
consecutive nights, at the start and very end of 2010 July 25. On the
first night the seeing was poor at about 1\farcs1, but this improved
to 0\farcs6 on the second night. We employed the same observing
strategy on both nights. A 9-point dither pattern, with three 10-s
exposures at each dither position, was repeated twice, and resulted in
a total exposure time of 540 s. As the source density in the target
field is very high, we also observed a relatively empty field centered
on an interstellar dark cloud that is 50\arcmin~away from
GX\,3+1. From these dithered offset sequences, taken immediately
following the GX\,3+1 observations, we constructed sky-background
maps.

We used the PANIC data-reduction package written for IRAF to reduce
the data. After dark-subtracting all science frames, the PANIC
pipeline averages the exposures taken at each dither position, and
corrects for non-linearity of the detector response. Flat-fielding is
achieved using master twilight flats. The processed offset-field
exposures are median-combined to create an initial sky map. Objects
detected in the sky-subtracted offset frames are masked out before
doing a second iteration of median-combining, which creates the final
sky map. The sky-subtracted GX\,3+1 exposures are subsequently
corrected for astrometric distortion, and finally aligned and stacked
into one master image for each night. We tied the astrometry of these
master images to the International Celestial Reference System (ICRS)
using about twenty unsaturated and relatively isolated 2MASS stars in
the field. Fitting for zero point, rotation angle, and scale factor
gives an astrometric solution with an r.m.s.~scatter of
$\sim$0\farcs06~in right ascension and declination; this is comparable
to the intrinsic astrometric accuracy of 2MASS positions.

As a result of the poor seeing, the GX\,3+1 images of epoch 1 were of
little use to find, and determine the properties of, candidate NIR
counterparts. Instrumental magnitudes for the image from epoch 2 were
extracted with {\tt DAOPHOT} point-spread-function (PSF) fitting
photometry, and are calibrated using the $K_s$ magnitudes of eleven
isolated and well-fitted 2MASS stars in the field. We derived a
constant magnitude offset by averaging the differences between the
instrumental and calibrated magnitudes of the comparison stars, which
gives an r.m.s.~scatter of 0.065 mag around the mean offset.

\subsubsection{FourStar imaging} \label{sec_fourstar}

On 2012 May 2 and August 8 we imaged the field of GX\,3+1 in the $K_s$
band with the FourStar camera \citep{monsea11} on the Magellan Baade
telescope. The FourStar $2\times2$ array of $2048\times2048$ pixel$^2$
HAWAII-2RG detectors has a 0\farcs159 pixel$^{-1}$ plate scale on the
Baade, and provides a full field of view of 10\farcm8 $\times$
10\farcm8 with 19\arcsec-wide gaps. We placed our target near the
center of one of the four detectors. The seeing on May 2 was excellent
($\sim$0\farcs4), but the poor seeing ($\sim$1\farcs1) during the
second run made the images collected on August 8 of little use for
achieving our science goals.

On May 2 we employed a 9-point dither pattern with fifteen 4.4-s
exposures at each dither position, for a total observing time of 594
s. Two sequences of an offset field bracketed the target
observations. On August 8 we took one target and one offset sequence,
each consisting of 9 dithers with nine 8.7-s exposures at each
position, for a total exposure of $\sim$705 s. The data reduction
steps are similar to those adopted for the PANIC images
(Sect.\ref{sec_panic}), except that for the case of the FourStar
images we used the {\tt SCAMP} package \citep{bert06} to correct for
the geometric distortion. To this end, fifth-order polynomials were
fitted to the catalogued and measured positions of 2MASS stars to map
out the variable plate scale over the chip area.  With the {\tt SWarp}
routines \citep{bertea02} the images were resampled to
distortion-corrected images with a linear plate scale of 0\farcs12
pixel$^{-1}$, similar to the PANIC plate scale. The small degree of
oversampling with respect to the intrinsic FourStar plate scale is
justified by the non-integer dither offsets of the observing
sequences. The astrometry of the final images was tied to the ICRS in
the same way as we did for the PANIC images, which resulted in an
astrometric solution with an r.m.s.~scatter of $\sim$0\farcs065~in
right ascension and declination.

PSF photometry was performed on the final stacked image of epoch 3. The
absolute calibration of the FourStar photometry was derived based on
the calibrated PANIC photometry. We chose eleven isolated stars within
11\arcsec~of GX\,3+1 with a clean PSF measurement in the PANIC and
FourStar images, and $14.3<K_s<16.0$ to derive the mean magnitude
offset that converts FourStar instrumental to calibrated
magnitudes. The r.m.s.~scatter around this mean offset is 0.034 mag.

\subsubsection{FIRE spectroscopy} \label{sec_fire}

NIR spectra of two candidate counterparts (A\,1 and A\,2; see
Sect.~\ref{sec_id}) were obtained with the Folded-port InfraRed
Echelette (FIRE; \citealt{simcea13}) spectrograph on the Magellan
Baade telescope.  FIRE is equipped with a $2048 \times 2048$ pixel$^2$
HAWAII-2RG HgCdTe detector. The spectrograph was used in the
low-resolution longslit prism mode with a slit width of 0\farcs6,
yielding a continuous coverage of the 0.8--2.5 \micron~band and a
resolving power in the $K_s$ band of $R=300$.  In our setup the
spatial sampling across the slit is 0\farcs15 per pixel.

Observations were taken on the night of 2012 May 3 starting at 09:02
UT when GX\,3+1 was at an airmass of $\sim$1.05. The slit was
positioned such that both candidate counterparts fell in the slit.
Observing conditions were good with a typical seeing of 0\farcs6; this
is good enough to separate the PSF peaks of A\,1 and A\,2, which are
only $\sim$0\farcs76 apart. We executed a sequence of nine 63.4-s
exposures, and nodded the targets along the slit.  The bright
($V=8.3$) telluric standard HD\,169291 of spectral type A0\,V was
observed after the GX\,3+1 observations with a sequence of five 1-s
exposures.

We reduced the data using mainly IRAF routines. A master flat was
created from two sets of dome flats, each optimized to have sufficient
counts in either the blue or red part of the spectrum. Next, science
exposures were subtracted in pairs; each frame was paired with the
exposure nearest in time in which the stars were dithered to an offset
position. This yields an initial sky subtraction, and enabled us to
locate the stars in the slit. The resulting difference images were
divided by the masterflat. In this crowded field there are hardly any
clean background patches to measure the residual sky. For that reason
we chose a few narrow background regions close to the targets, but
especially in the $H$ band ($\sim$1.5--1.8 $\micron$), where the sky
OH emission lines are strong, the signatures of an imperfect sky
subtraction can be seen as sharp features in the final spectra, which
add to the noise. We extracted the spectra by simply adding the
sky-subtracted pixel values within the aperture limits.  Given the
small angular separation between the two targets, we defined a narrow,
2-pixel wide, extraction aperture to minimize the stars' mutual
contamination but as a result of our extraction method, it cannot be
removed completely. The widely-used optimal-extraction algorithm by
\cite{horn} was developed for isolated stars, and indeed did not
produce good results for the blended spatial profiles of our
targets. The wavelength calibration was done using a Neon-Argon lamp
exposure taken immediately following the GX\,3+1 sequence. The errors
in the dispersion solution are $\lesssim$0.001 $\micron$. Redward of
2.1 $\micron$ there are only a few arc lines and a rapidly-rising
thermal background from the telescope; this makes the wavelength
calibration very uncertain in that part of the spectrum. The
individual spectra of each star were average-combined into a master
spectrum after scaling the relatively featureless stretch of continuum
between 2.1 and 2.15 $\micron$ to the same value. A master spectrum
for the telluric standard was extracted in a similar way. We performed
the telluric correction using the method and routines described in
\cite{vaccea03}.

\begin{figure}
\centerline{\includegraphics[width=8.5cm]{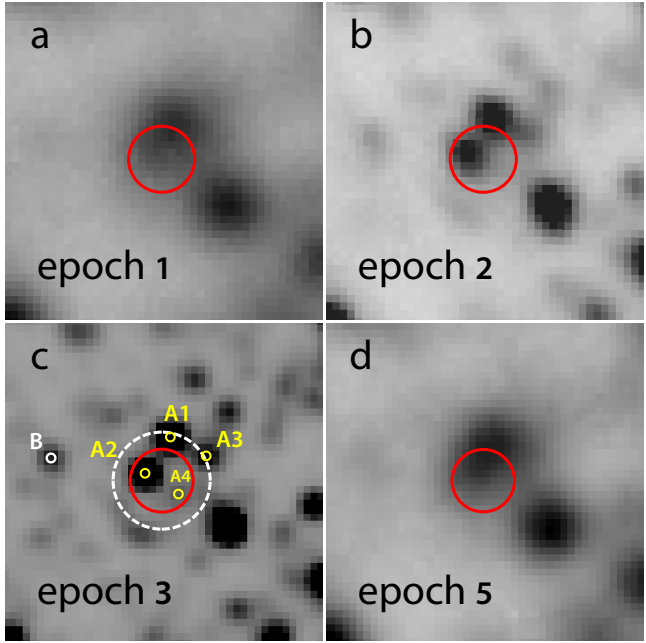}}
\caption{PANIC (top) and FourStar (bottom) images of all four imaging
  epochs centered on the new {\em Chandra} position of GX\,3+1. The
  90\% confidence errors ($\sim$0\farcs58) are indicated with red
  circles, whereas the 95\% confidence error ($\sim$0\farcs83) is
  shown as a white dashed circle in panel {\em c} only. Small yellow
  circles mark the positions of candidate NIR counterparts. Source B
  was considered to be a possible counterpart by \cite{zolorevn11} but
  this is ruled out by our new position. Each image is about 5\farcs5
  $\times$ 5\farcs5~in size.  North is up, east to the
  left.\label{fig_nir}}
\end{figure}

\section{The NIR counterpart to GX\,3+1} \label{sec_id}

Figure \ref{fig_nir} shows the region around the new {\em Chandra}
position of GX\,3+1 in the $K_s$ images from all four imaging
epochs. The red circles represent the 90\% confidence radius on the
source position. These were computed by adding in quadrature the {\em
  Chandra} absolute pointing error (0\farcs54) and the errors in the
astrometric solutions of the PANIC and FourStar images (see
Sections~\ref{sec_panic} and \ref{sec_fourstar}) scaled to 90\% errors
assuming a 2-D Gaussian distribution. For illustrative purposes we
also show the combined 95\% error radius in
Fig.~\ref{fig_nir}c\footnote{We note that the 95\% confidence limit on
  the absolute pointing of the HRC-I is not known very accurately. We
  conservatively assumed 0\farcs8; see the link in the previous
  footnote.}. From Fig.~\ref{fig_nir}c, which shows the image taken
under the best seeing conditions, it is clear that multiple $K_s$-band
sources are viable counterparts to GX\,3+1 based on their locations
inside or very close to the 90\% (0\farcs58 for this epoch) or 95\%
(0\farcs83) confidence radii; in principle, any of the sources marked
with a small yellow circle could be the counterpart. Comparison of
this image with the lower-resolution UKIRT $K$ image in Fig.~5 of
\cite{zolorevn11} shows that their candidate counterpart A is now
resolved into four sources, which we label A\,1 to A\,4 in order of
decreasing $K_s$ brightness. Source B, the other counterpart proposed
by \cite{zolorevn11}, lies too far from the new {\em Chandra} position
to still be considered a possible counterpart.  Table~\ref{tab_ids}
lists the coordinates of these stars together with their $K_s$
magnitudes from epochs 2 and 3, during which they are detected without
significantly suffering from blending.

\begin{figure*}
\centerline{\includegraphics[width=18cm]{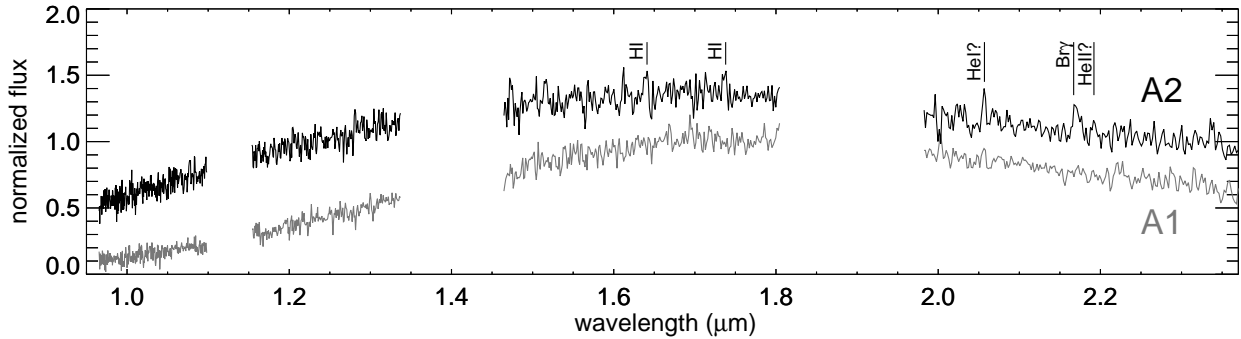}}
\caption{FIRE spectra of stars A\,1 (bottom spectrum, gray) and A\,2
  (top, black). Emission lines of H\,I, and possibly He\,I and He\,II,
  that are (tentatively) identified in the spectrum of A\,2 are
  marked. Sharp features in the spectrum, mostly apparent in the
  $H$ band, result from imperfect sky subtraction. The spectra are
  normalized to the flux at 1.66 \micron~and an arbitrary offset of
  0.3 flux units is applied to the spectrum of A\,2 for clarity. Parts
  of the spectra where the transmission through the Earth's atmosphere
  is low are not plotted. \label{fig_nirspec}}
\end{figure*}

The two candidate counterparts for which we obtained FIRE spectra are
A\,1 and A\,2, which lie just outside and inside the 90\% error
circle, respectively. Whereas the spectrum of A\,1 is featureless, the
spectrum of A\,2 clearly shows Br\,$\gamma$ in emission
(Figure~\ref{fig_nirspec}). This H\,I emission feature is commonly
associated with accreting sources \citep{bandea99,bandea03}, and its
presence shows that A\,2 is the true NIR counterpart of GX\,3+1. Other
H\,I, and possibly He\,I and He\,II, lines may be identified as well
but are much weaker; especially the 2.192 \micron~feature is very weak
and may turn out not to be a real emission line in a higher
signal-to-noise spectrum. Nevertheless, the spectrum of A\,1 does not
show similar features at the corresponding wavelengths. In
Table~\ref{tab_lines} we list the measured wavelengths of the
(tentative) emission lines. The equivalent width of the Br\,$\gamma$
line is about $-45\pm5$ \AA; given that the spectrum of A\,2 still
contains a small contribution from the light of A\,1, the true value
must be lower (i.e.~more negative). No absorption features from a
companion star can be seen in the spectrum of A\,2 but we point out
the poor signal-to-noise ratio especially in the $H$ band (due to poor
sky subtraction) and beyond $\sim$2.3 $\micron$, where the CO
absorption bands appear as prominent features in the $K$-band spectra
of stars of spectral type K and M.

We compare the magnitudes for star A\,2 from epochs 2 and 3 to check
for variability.  Taking into account both the formal photometry
errors from {\tt DAOPHOT} and the photometric calibration errors, we
do not see any sign of variability in $K_s$ in excess of $\sim$0.1 mag
in our own observations (Table~\ref{tab_ids}). The combined magnitudes
of A\,1 to A\,4 are consistent with $K=14.87\pm0.13$ mag for source A
on 2007 May 3 as reported by \cite{zolorevn11}, and with
$K=15.13\pm0.16$ mag for that same source (i.e.~star 311) on 1988 May
11--15 as reported by \cite{naylea91}. This points to a lack of
large-amplitude NIR variability on a time scale of years. 

\begin{figure}
\centerline{\includegraphics[width=9.5cm]{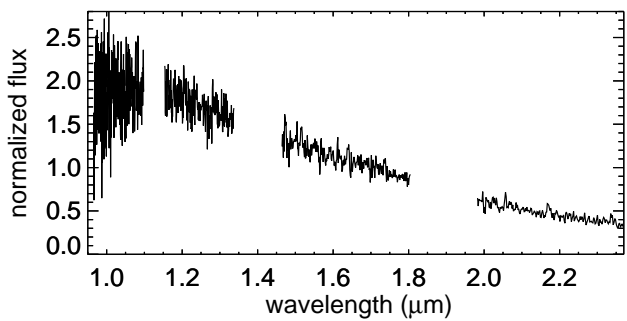}}
\caption{Dereddened spectrum of A\,2 normalized to the flux at 1.66
  \micron. We adopted $N_H=1.59 \times 10^{22}$ cm$^{-2}$ following
  \cite{oostea01} and assumed the extinction law from
  \cite{cardea89}. Adopting the extinction coefficients from
  \cite{nishea08} would result in a steeper continuum slope.
   \label{fig_nirspec_dered}}
\end{figure}

\begin{table}
\caption{Positions and magnitudes of NIR sources near the new
  position of GX\,3+1 \label{tab_ids}}
\begin{center}
\begin{tabular}{c@{\hskip0.2cm}c@{\hskip0.2cm}c@{\hskip0.2cm}c@{\hskip0.2cm}c@{\hskip0.2cm}c}
\hline
\hline
ID   & $\alpha_{\rm J2000}$ & $\delta_{\rm J2000}$ & $\Delta$ (\arcsec) & $K_{s,{\rm epoch 2}}$ & $K_{s,{\rm epoch 3}}$ \\   
\hline
A\,1   & 17$^{\rm h}$47$^{\rm m}$56.063$^{\rm s}$ & $-$26$^{\circ}$33\arcmin48.74\arcsec  & 0.76 & 15.62$\pm$0.01 & 15.62$\pm$0.01\\
A\,2   & 17$^{\rm h}$47$^{\rm m}$56.097$^{\rm s}$ & $-$26$^{\circ}$33\arcmin49.35\arcsec  & 0.31 & 15.82$\pm$0.01 & 15.84$\pm$0.01\\
A\,3   & 17$^{\rm h}$47$^{\rm m}$56.015$^{\rm s}$ & $-$26$^{\circ}$33\arcmin49.06\arcsec  & 0.87 & 16.78$\pm$0.02 & 16.84$\pm$0.02\\
A\,4   & 17$^{\rm h}$47$^{\rm m}$56.056$^{\rm s}$ & $-$26$^{\circ}$33\arcmin49.71\arcsec  & 0.37 & 18.12$\pm$0.05 & 18.11$\pm$0.04\\
B      & 17$^{\rm h}$47$^{\rm m}$56.214$^{\rm s}$ & $-$26$^{\circ}$33\arcmin49.09\arcsec  & 1.94 & 17.26$\pm$0.02 & 17.13$\pm$0.02\\
\hline
\end{tabular}
\end{center}
\tablecomments{$\Delta$ is the offset between the positions of the NIR
  sources and the new {\em Chandra} position of GX\,3+1. We consider
  A\,2 to be the likely NIR counterpart to GX\,3+1 based on the
  detection of a Br\,$\gamma$ emission line in its $K_s$-band
  spectrum. The uncertainties in the magnitudes are the {\tt DAOPHOT}
  errors on the PSF photometry. Additional errors in the photometric
  calibration with respect to 2MASS are $<$0.1 mag and are given in
  Sections~\ref{sec_panic} and \ref{sec_fourstar}. }
\end{table}

\begin{table}
\begin{center}
\caption{Emission lines in the spectrum of A\,2 \label{tab_lines}}
\begin{tabular}{lccc}
\hline
\hline
Line         & $\lambda_{\rm obs}$ & $\lambda_{\rm vacuum}$ & Equivalent width \\
             & ($\micron$)       & ($\micron$)          & (\AA) \\
\hline
H\,I         & 1.641             & 1.642 & $-$17$\pm$3 \\
H\,I         & 1.738             & 1.737 & $-$13$\pm$3 \\
He\,I?       & 2.057             & 2.059 & $-$13$\pm$3 \\
Br\,$\gamma$ & 2.167             & 2.166 & $-$45$\pm$5 \\
He\,II?      & 2.192             & 2.189 & $-$19$\pm$3 \\
\hline
\end{tabular}
\tablecomments{Observed wavelengths $\lambda_{\rm obs}$ correspond to
  the locations of the peak flux values in the lines. Errors in the
  wavelengths are about $\pm$0.001 $\micron$. Due to the contamination
  by the light of A\,1, the equivelent-width values are upper limits.}
\end{center}
\end{table}

We also searched the Vista Variables in the Via Lactea (VVV) catalogs
\citep{minnea10} for detections of A\,2. The most recent data release
(DR), i.e.~DR\,3, has 18 images that cover the position of GX\,3+1,
including 10 epochs in the $K_s$ band. However, in these images A\,2
is not detected as a separate object but as a blend with, at least,
A\,1, or with A\,1 and the bright object to the south-west of A\,4
just outside the 95\% error circle (see Fig.~\ref{fig_nir}c). The
poorer image quality of the VVV images compared to that of our images
from epochs 2 and 3 is due partly to the coarser VVV pixel scale
(0\farcs34 pixel$^{-1}$) and partly to the seeing conditions under
which these VVV images were taken ($\gtrsim$0\farcs8). Reprocessing of
the VVV images to attempt to extract deblended magnitudes for A\,2 and
A\,1 is beyond the scope of this paper. A visual examination of the
DR\,3 images did not reveal any obvious brightness variations of
A\,2. The recently-released multi-band master catalog extracted from
the DR\,1
images\footnote{http://www.eso.org/sci/observing/phase3/data\_releases/VVV\_CAT.2014-07-11.pdf}
did not include a detection of A\,2, either.

\section{Discussion and conclusions} \label{sec_disc}

The NIR emission from bright NS-LMXBs can originate in several parts
of the system: the accretion disk may contribute via thermal emission
that results from X-ray or viscous heating, a late-type secondary can
produce thermal emission in the NIR which may be enhanced by X-ray
heating of the hemisphere facing the accretion region, and finally the
inner region of a jet---if present---may contribute through
optically-thin synchrotron emission. {\em Spitzer} data of the NS-LMXB
4U\,0614+091 suggest that one would have to observe at
wavelengths $>$8 $\micron$ to detect emission from the optically-thick
part of the jet \citep{miglea10}. For jets that are significantly
  stronger than the one in 4U\,0614+091 the break between the
  optically-thin and thick part of the jet may shift to shorter
  wavelengths \citep{falcea04}.

The FIRE spectrum of GX\,3+1 indicates that at least a significant
portion of the NIR light comes from a heated accretion disk or heated
secondary. Both the emission lines and the blue (i.e.~$F_{\nu} \propto
\nu^{\alpha}$ with $\alpha > 0$) continuum of the unreddened spectrum
(Figure~\ref{fig_nirspec_dered}) are as expected for a thermal
component \citep{russea07}. Based on our NIR data alone we are not
able to distinguish between thermal disk emission resulting from
either X-ray or viscous heating. This would require simultaneous NIR
and X-ray observations to look for possible correlated behavior on
time scales of seconds, as would be predicted by X-ray heating.

GX\,3+1 has not been detected at radio wavelengths \citep{bereea00},
suggesting there is no powerful jet in this system. Furthermore,
optically-thin synchrotron emission produces a red ($\alpha < 0$) NIR
spectrum, which clearly is inconsistent with the spectrum in
Figure~\ref{fig_nirspec_dered}. We note that the exact slope of the
continuum is uncertain as contamination by the light from the close
neighbor A\,1 is not completely accounted for (see
Sect.~\ref{sec_fire}). However, in Figure~\ref{fig_nirspec} one can
see that the observed spectrum of A\,2 is bluer than that of A\,1, so,
if anything, the effect of A\,1 is to make the spectrum of A\,2 seem
redder. We conclude that most likely a jet does not significantly
contribute to the NIR emission of GX\,3+1.

To estimate the possible contribution of an unheated secondary, we
first use the estimate of the column density towards GX\,3+1 from
\cite{oostea01}, viz.~$N_H=1.59^{+0.07}_{-0.12}\times10^{22}$
cm$^{-2}$, and the distance of 6.1 kpc (with an estimated uncertainty
of $\sim$15\%, \cite{kuulea03}) to compute the absolute $K_s$
magnitude of A\,2. Adopting the relation $N_H=1.79 \times
10^{21}\,A_{V}$ cm$^{-2}$ from \cite{predschm95}, $A_K=0.114 \times
A_{V}$ from \cite{cardea89}, and $A_K=0.95 \times A_{K_s}$ from
\cite{dutrea02}, we find $M_{K_s}=0.84\pm0.35$, where the error is
dominated by the uncertainty in the distance. However, since GX\,3+1
lies close to the direction of the Galactic Center, it may be more
appropriate to adopt the conversion $A_{K_s}=0.062 \times A_{V}$ from
\cite{nishea08}, which yields $M_{K_s}=1.35\pm0.34$.  We compare this
value to the absolute $K_s$ magnitudes of late-type dwarfs and giants,
which typically are the donors in LMXBs. Main-sequence stars of
spectral type G0 to M5 have $M_{K_s} = 3.0$ to 8.4 (Mamajek
2014\footnote{http://www.pas.rochester.edu/$\sim$emamajek/EEM\_dwarf\_UBVIJHK\_
  \mbox{colors}\_Teff.txt}). Therefore, if the secondary is indeed a
late-type dwarf, it does not provide a dominant ($>50\%$) contribution
to the $K_s$ flux. Giants of spectral type G0 and later have
$M_{K_s}\lesssim -0.45$ \citep{ostlcarr07,bessbret88} which is
brighter than the bright limit to the estimated $M_{K_s}$ for GX\,3+1;
therefore we can exclude that the secondary is a late-type giant.

Given that the $K_s$ flux from GX\,3+1 is most likely thermal, we can
estimate the orbital period $P_{b}$ using $M_{K_s}$ and the X-ray
luminosity $L_X$. A relation between these properties is expected if
the NIR emission originates from thermal emission in the outer parts
of the accretion disk (where X-ray reprocessing dominates over viscous
heating), the size of which is set by the orbital separation of the
binary stars---and thus the orbital period. \cite{vanpmccl94}
demonstrated the existence of such a correlation in the optical, and
it has recently been extended to the NIR by \cite{revnea12} based on a
small sample of NS-LMXBs. For $L_X/L_{\rm Edd} = 0.1-0.2$ and
$M_{K_s}=1.35\pm0.34$ we find that $P_b \approx 4.4-10.9$ h using the
equation in Sect.~4.2 in \cite{revnea12}\footnote{\cite{revnea12} use
  the 2--10 keV luminosity rather than the bolometric luminosity,
  which is why the adopted $L_X/L_{\rm Edd}$ range is 0.1--0.2
  (reflecting the $(2-4)\times10^{37}$ erg s$^{-1}$ range found by
  \cite{denhea03}, see Sect.\ref{sec_intro}) rather than 0.15--0.3.};
for $M_{K_s}=0.84\pm0.35$ we find $P_b \approx 7.3-18.5$ h. This range
of periods also suggests the secondary is not a giant \cite[see
  e.g.][]{verb93}, although it leaves room for a somewhat evolved star
or sub-giant. Now that we have identified the NIR counterpart of
GX\,3+1, a targeted follow-up campaign of time-resolved spectroscopy
or photometry may be attempted to measure the orbital period directly.

\cite{russea07} investigated the origin of the OIR emission in
NS-LMXBs based on a set of near-simultaneous X-ray and OIR data. They
found that both at luminosities below $L_X\approx10^{36}$ erg s$^{-1}$
and at the high X-ray--luminosity end (i.e.\,in the Z sources) most of
the NIR emission comes from an X-ray--heated disk; for atolls and
milli-second X-ray pulsars of intermediate luminosity the jet emission
dominates. The sample of NS-LMXBs studied by Russell et al.~did not
include the brightest atolls. In fact, NIR counterparts were
discovered only recently for several of these systems like GX\,9+1
\citep{currea11}, 4U\,1705$-$44 \citep{homaea09}, and now GX\,3+1
(this work). The counterpart of GX\,9+1 has not been studied in
detail, yet. From $JHK_s$ colors and correlations between
near-simultaneous X-ray (3--100 keV) and $K_s$-band data,
\cite{homaea09} found that in 4U\,1705$-$44, a source that can reach
similar X-ray luminosities as GX\,3+1, the NIR flux is likely the
result of X-ray heating, like in the Z sources. They suggested that
the high X-ray luminosity could be a sign of a larger disk compared to
the disks in fainter atolls, which would explain the larger NIR
contribution from X-ray heating. Our findings indicate that GX\,3+1
behaves similarly to 4U\,1705$-$44, in the sense that jet NIR emission
does not play a (dominant) role; therefore this scenario could offer a
plausible explanation for the origin of the NIR emission in GX\,3+1,
as well.

Finally, we comment on the lack of variability of the NIR counterpart
to GX\,3+1. In Figure~\ref{fig_xlightc} the quasi-periodic modulations
in the {\em RXTE}/ASM and {\em MAXI} count rate for GX\,3+1 are
clearly visible. If these brightness variations are truly the result
of changes in the mass-accretion rate, it is expected that the NIR
emission, which as we show contains a dominant contribution from an
accretion disk or heated secondary, should also vary in time. For
example, in the case of 4U\,1705-44 \cite{homaea09} clearly see
correlated X-ray and NIR variations. The difference in X-ray count
rate between the epochs of our imaging observations is a factor of
$\sim$2, but surprisingly, we see no change in the $K_s$ magnitude
within the photometric accuracy ($\sim$0.1 mag). We note that we only
have good measurements of the NIR brightness at two epochs, and need
further monitoring to investigate the (lack of) NIR variability of
GX\,3+1 in more detail.

\begin{acknowledgments}
The authors would like to thank A.~Monson for help with the FourStar
data reduction, and R.~Remillard, P.~Sullivan and M.~Matejek for
obtaining part of the observations. This work is supported by {\em
  Chandra} grant GO2-13048X.
\end{acknowledgments}

{\it Facilities:} \facility{CXO}, \facility{RXTE}, \facility{MAXI}, \facility{Magellan:Baade (PANIC, FourStar, FIRE)}

\end{document}